\documentclass[12pt,preprint,notoc,nohyper]{QCEMDA} 

\usepackage{epsfig,multicol,bbm}

\newcommand\fverb{\setbox\pippobox=\hbox\bgroup\verb}
\newcommand\fverbdo{\egroup\medskip\noindent
            \fbox{\unhbox\pippobox}\ }
\newcommand\fverbit{\egroup\item[\fbox{\unhbox\pippobox}]}
\newbox\pippobox

\title{Thermodynamics of charged and rotating black strings}

\author{Aeeman Fatima and K. Saifullah  \\

Department of Mathematics, Quaid-i-Azam University, Islamabad,
Pakistan \\

Electronic address: \email{saifullah@qau.edu.pk}}

\preprint{}  

\abstract{We study thermodynamics of cylindrically symmetric black
holes. Uncharged as well as charged and rotating objects have been
discussed. We derive surface gravity and hence the Hawking
temperature and entropy for all these cases. We correct some results
in the literature and present new ones. It is seen that
thermodynamically these black configurations behave differently from
spherically symmetric objects.}

\begin{document}

\section{Black strings}

The theory of thermal radiation and evaporation of black holes
\cite{Hawk1, Hawk2} was a breakthrough in theoretical physics, and
it became a subject where three different areas - the quantum
theory, general relativity and thermodynamics - converge. The laws
of black hole mechanics were established \cite{BCH, Ha} which are
analogous to the laws of ordinary thermodynamics. This also made
possible to apply physical investigations to these queer objects.
Initially only spherically symmetric black holes were investigated.
Later, spherical black holes with an axis of symmetry, like the Kerr
black hole were also studied. The study of prolate gravitational
collapse of cylindrical and other similar objects resulted in the
formulation of the hoop conjecture which states that horizons form
if and only if a mass gets compressed into a region whose
circumference in every direction is less than its Schwarzschild
circumference, $4\pi GM$. This excluded the possibility of the
formation of a cylindrical black hole. However, the hoop conjecture
holds only when the cosmological constant vanishes. Thus in the
presence of negative cosmological constant cylindrically symmetric
black holes (black strings) have been studied with great interest.
These represent asymptotically anti-de Sitter spaces in the
transverse and axial directions. After the pioneering work
\cite{lemos95a,lemos95b, LZ, S, CZ} on their fundamental structure
and properties, these interesting objects have been investigated in
other contexts as well. They have been studied, for example, in the
presence of Born-Infeld and Maxwell fields, in higher dimensional
and Gauss-Bonnet gravity theories \cite{BCGZ, CBSV, He, CPTZ}, and
in the framework of supergravity theories, low energy string
theories and topological defects \cite{HT, AK, Ka}. Their study is
significant from the point of view of cosmic strings and toroidal
black holes as well which look locally like black strings. In this
paper we study thermodynamic properties of uncharged, charged and
rotating black strings. Apart from recovering results that already
exist, we correct some formulae in the literature and present new
ones as well.

We consider the Einstein-Hilbert action in the presence of the
cosmological constant and an electromagnetic field. The total action
is \cite{LZ}
\begin{equation}
\mathcal{S}+\mathcal{S}_{em}=\frac{1}{16\pi G}\int
d^{4}x\sqrt{-g}\left( R-2\Lambda \right) - \frac{1}{16\pi }\int
d^{4}x\sqrt{-g}F^{\mu \nu }F_{\mu \nu }, \label{2.8}
\end{equation}
where $\mathcal{S}$ is given by
\begin{equation}
\mathcal{S}=\frac{1}{16\pi G}\int d^{4}x\sqrt{-g}\left( R-2\Lambda
\right) , \label{2.9}
\end{equation}
and $R$ is the curvature scalar, $g$ is the determinant of the
metric. The Maxwell tensor is given as
\begin{equation}
F_{\mu \nu }=\partial _{\mu }A_{\nu }-\partial _{\nu }A_{\mu },
\label{2.10}
\end{equation}%
where $A_{\mu }$ being the vector potential is given by
\begin{equation}
A_{\mu }=-h(r)\delta _{\mu }^{0},  \label{2.11}
\end{equation}%
$h\left( r\right) $ is an arbitrary function of the radial
coordinate $r$. In this paper we study the solutions of the
Einstein-Maxwell equations in cylindrical coordinates $\left(
t,r,\phi ,z\right)$ with
\begin{equation}
-\infty <t<\infty , 0\leq r<\infty , 0\leq \phi \leq 2\pi , -\infty
<z<\infty .  \label{2.12}
\end{equation}

We will give explicit form of metrics for the uncharged, charged and
rotating cases in subsequent sections.

\section{Thermodynamics of uncharged black string}

The general form of the static uncharged black string metric with
negative cosmological constant, $\alpha ^{2}=-\frac{1}{3}\Lambda
>0$, in asymptotically anti-de Sitter direction has been constructed as
\cite{CZ}
\begin{equation}
ds^{2}=-\left( \alpha ^{2}r^{2}-\frac{4M}{\alpha r}\right)
dt^{2}+\left( \alpha ^{2}r^{2}-\frac{4M}{\alpha r}\right) ^{-1}
dr^{2}+r^{2}d\phi ^{2}+\alpha ^{2}r^{2}dz^{2},  \label{3.1}
\end{equation}%
where $\alpha =-\Lambda /3$, $\Lambda$ being the cosmological
constant, and $M$ is related to the ADM mass density of the black
string.

Putting $g^{11}=0$ gives

\begin{equation}
\alpha ^{2}r^{2}-\frac{4M}{\alpha r}=0,   \label{3.3}
\end{equation}
so that the event horizon of the black string is
\begin{equation}
r_{+}=\frac{\left( 4M\right) ^{\frac{1}{3}}}{\alpha }.  \label{3.10}
\end{equation}

In order to find the surface gravity, we note that for a general
static and cylindrical symmetric metric of the form
\begin{equation}
ds^{2}=-f(r)dt^{2}+\frac{dr^{2}}{g(r)}+r^{2}d\phi ^{2}+\alpha
^{2}r^{2}dz^{2},        \label{3.11}
\end{equation}
there is a coordinate singularity at $r=r_{+}$ i.e. $g(r_{+})=0$
which can be removed by using the Painlev\`{e}-type coordinate
transformation
\begin{equation}
dt\rightarrow dt-\sqrt{\frac{1-g}{f g}}dr,   \label{3.12}
\end{equation}
so that the metric becomes
\begin{equation}
ds^{2}=-fdt^{2}+dr^{2}+2f\sqrt{\frac{1-g}{f g}}drdt+r^{2}d\phi
^{2}+\alpha^{2}r^{2}dz^{2}.  \label{3.13}
\end{equation}
Near the event horizon we use Taylor's series to expand the
functions, $f$ and $g$ as
\begin{equation}
f(r_{+})=f^{\prime }(r_{+})(r-r_{+})+O((r-r_{+})^{2} , \label{3.14}
\end{equation}
\begin{equation}
g(r_{+})=g^{\prime }(r_{+})(r-r_{+})+O((r-r_{+})^{2} . \label{3.15}
\end{equation}
In this form the surface gravity becomes

\begin{equation}
\kappa =\left\vert \Gamma _{00}^{0}\right\vert
_{r=r_{+}}=\frac{1}{2} \left\vert
\sqrt{\frac{1-g}{fg}}g\frac{df}{dr} \right\vert _{r=r_{+}}.
\label{3.16}
\end{equation}
Using this formula for the black string (\ref{3.1}) we obtain
\begin{eqnarray}
\kappa &=&\frac{\alpha ^{3}r_{+}^{3}+4M}{\alpha r_{+}^{2}}.
\label{3.17}
\end{eqnarray}%
Using Eq. (\ref{3.10}) in Eq. (\ref{3.17}) the surface gravity
becomes
\begin{eqnarray}
\kappa &=&3\alpha \left( \frac{M}{2}\right) ^{\frac{1}{3}}.
\label{3.18}
\end{eqnarray}
The Hawking temperature, $T=\kappa /2\pi$, of uncharged black string
is thus given by

\begin{equation}
T=\frac{3\alpha }{2\pi }\left( \frac{M}{2}\right) ^{\frac{1}{3}},
\label{3.20}
\end{equation}
The temperature goes as $M^{\frac{1}{3}}$, which is very different
from that of the Schwarzscild black hole \cite{2}. Thus the negative
cosmological constant and topological difference change the
thermodynamical behaviour of black configurations.

The area of the event horizon \cite{CZ} of a black string, $\sigma
=2\pi \alpha r_{+}^{2}$, in this case becomes
\begin{equation}
\sigma =\frac{2\pi \left( 4M\right) ^{\frac{2}{3}}}{\alpha }.
\label{3.22}
\end{equation}
Thus the Bekenstein-Hawking entropy relation in geometrized units,
$S=\sigma/4$, takes the form
\begin{equation}
S=\frac{\pi }{2\alpha }\left( 4M\right) ^{\frac{2}{3}}.
\label{3.24}
\end{equation}

\section{Thermodynamics of charged black string}

In this section we discuss thermodynamics of static charged black
string. The spacetime in asymptotically anti-de-Sitter direction can
be written as \cite{CZ}
\begin{equation}
ds^{2}=-\left( \alpha ^{2}r^{2}-\frac{4M}{\alpha
r}+\frac{4Q^{2}}{\alpha
^{2}r^{2}}\right) dt^{2}+\left( \alpha ^{2}r^{2}-\frac{4M}{\alpha r}+\frac{%
4Q^{2}}{\alpha ^{2}r^{2}}\right) ^{-1}dr^{2}+r^{2}d\phi ^{2}+\alpha
^{2}r^{2}dz^{2},  \label{3.26}
\end{equation}
where $M$ and $Q$ are constants. Note that for a cylinder of
infinite radius and height $\Delta z$ the total charge will be
infinite. However, the line charge density $Q_{z}/\Delta z$ will be
finite. We take this quantity as $Q$ and it is related to ADM
charge. Similarly $M$ is the ADM mass per unit length in the $z$
direction.

The event horizon can be found by putting $g^{11}=0$ as
\begin{equation}
\alpha ^{2}r^{2}-\frac{4M}{\alpha r}+\frac{4Q^{2}}{\alpha
^{2}r^{2}}=0.  \label{3.27}
\end{equation}
Solving this quartic equation and discarding the two imaginary
roots, we see that the horizons of charged black string are
\begin{equation}
r_{\pm }=\frac{\left( 4M\right) ^{\frac{1}{3}}}{2\alpha }\left[
\sqrt{s}\pm \sqrt{2\sqrt{s^{2}-Q^{2}\left( \frac{2}{M}\right)
^{\frac{4}{3}}}-s}\right] ,
\end{equation}
where $s$ is given by

\begin{equation}
s=  \left( \frac{1}{2}+\frac{1}{2}\sqrt{1-\frac{64Q^{6}}{27M^{4}}}
\right) ^{\frac{1}{3}}+\left(
\frac{1}{2}-\frac{1}{2}\sqrt{1-\frac{64Q^{6}}{ 27M^{4}}}\right)
^{\frac{1}{3}}  .  \label{3.39}
\end{equation}
We can also express it in another form as
\begin{equation}
r_{\pm }=\frac{M^{\frac{1}{3}}}{2^{\frac{1}{3}}\alpha }\left[
\sqrt{x^{\frac{1}{3}}+y^{\frac{1}{3}}}+\sqrt{
-x^{\frac{1}{3}}-y^{\frac{1}{3}}+2\sqrt{-z+ \left[
x^{\frac{1}{3}}+y^{\frac{1}{3}}\right]^{2}}}\right] , \label{3.41}
\end{equation}
where
\begin{equation}
x=\frac{1}{2}+\frac{1}{2}\sqrt{1-\left( \frac{2}{M}\right) ^{
\frac{4}{3}}Q^{2}} {, \ \ \ }y=\frac{1}{2}-\frac{1}{2}\sqrt{1-\left(
\frac{2}{M}\right) ^{\frac{4}{3}}Q^{2}} , z=\left(
\frac{2}{M}\right) ^{\frac{4}{3}}Q^{2}.
\end{equation}
As $g^{11}\geq 0$ when $0\leq r\leq r_{-}$ and $r\geq r_{+}$;
$g^{11}\leq 0$ when $r_{-}$ $\leq r\leq r_{+}$, therefore, the two
positive roots can be taken as the outer and inner horizons of the
black string. The singularity at $r=0$ is enclosed by event
horizons. We note that Eq. (\ref{3.27}) has four solutions: $z=\pm
r_{+}$ give two outer horizons, and $z=\pm r_{-}$ two inner
horizons. These enclose the singularity at $z=0$.

Now, the surface gravity of a cylindrical metric (\ref{3.11}) can be
evaluated from \cite{17}
\begin{equation}
\kappa =\frac{1}{2\sqrt{-h}} \frac{\partial }{\partial x^{a}}\left(
\sqrt{-h}h^{ab}\frac{\partial r}{\partial x^{b}}\right) ,
\label{3.43}
\end{equation}
where the second order diagonal metric $h_{ab}$ and its inverse
$h^{ab}$, are from the $t-r$ sector of the metric. Substituting the
values we get the surface gravity as
\begin{eqnarray}
\kappa &=&\alpha ^{2}r_{+}+\frac{2M}{\alpha
r_{+}^{2}}-\frac{4Q^{2}}{\alpha ^{2}r_{+}^{3}},  \label{3.50}
\end{eqnarray}
or using the value of $r_+$ this takes the form
\begin{equation}
\kappa =\frac{2\alpha \left[
(M/2)^{\frac{4}{3}}D^{4}+(4M^4)^{\frac{1 }{3}}D-4Q^{2}\right]
}{MD^{3}},  \label{3.51}
\end{equation}
where
\begin{equation}
D=  \sqrt{x^{\frac{1}{3}}+y^{\frac{1}{3}}}
+\sqrt{-x^{\frac{1}{3}}-y^{\frac{1}{3}}+2\sqrt{ -z+\left[
x^{\frac{1}{3}}+y^{\frac{1}{3}} \right] ^{2}}}  . \label{3.52}
\end{equation}
Thus the Hawking temperature of charged black string becomes

\begin{equation}
T=\frac{\alpha \left[ (M/2)^{\frac{4}{3}}D^{4}+(4M^4)^{\frac{1
}{3}}D-4Q^{2}\right] }{\pi MD^{3}}.  \label{3.53}
\end{equation}
The area of the horizon per unit length in this case  takes the form

\begin{equation}
\sigma =\frac{\pi (2M^2)^{\frac{1}{3}}D^{2}}{\alpha }, \label{3.54}
\end{equation}
and thus the Bekenstein-Hawking entropy, $S=\sigma/4$, becomes
\begin{equation}
S=\frac{\pi (2M^2)^{\frac{1}{3}}D^{2}}{4 \alpha }. \label{3.57}
\end{equation}
In these terms the electric potential, $\Phi = 2Q/\alpha r_{+}$, of
the charged black string is given by
\begin{equation}
\Phi =\left(\frac{16}{M}\right)^{\frac{1}{3}}\frac{Q}{D}.
\label{3.56}
\end{equation}

\section{Thermodynamics of charged rotating black string}

In this section, we extend our study of thermodynamics to charged
and rotating black string. This is the most general case and other
cases can be derived from this one. The most general form of the
cylindrical symmetric spacetime asymptotically anti-de Sitter is
given as \cite{LZ}
\begin{eqnarray}
ds^{2} &=&-\left( \alpha ^{2}r^{2}-\frac{4M\left( 1-\frac{\alpha
^{2}a^{2}}{2 }\right) }{\alpha r}+\frac{4Q^{2}}{\alpha
^{2}r^{2}}\right) dt^{2} \nonumber  \\ &&-\frac{4aM
\sqrt{1-\frac{\alpha ^{2}a^{2}}{2}}}{\alpha r}\left(
1-\frac{Q^{2}}{M\left(
1-\frac{\alpha ^{2}a^{2}}{2}\right) \alpha r}\right) 2dtd\phi  \nonumber  \\
&&+\left( \alpha ^{2}r^{2}-\frac{4M\left( 1-\frac{3\alpha
^{2}a^{2}}{2} \right) }{\alpha r}+\frac{4Q^{2}}{\alpha
^{2}r^{2}}\frac{\left( 1-\frac{ 3\alpha ^{2}a^{2}}{2}\right)
}{\left( 1-\frac{\alpha ^{2}a^{2}}{2}\right) }
\right) ^{-1}dr^{2}  \nonumber  \\
&&+\left[ r^{2}+\frac{4Ma^{2}}{\alpha r}\left( 1-\frac{Q^{2}}{\left(
1-\frac{ \alpha ^{2}a^{2}}{2}\right) M\alpha r}\right) \right] d\phi
^{2}+\alpha ^{2}r^{2}dz^{2}.  \label{4.40}
\end{eqnarray}
As mentioned earlier the two constants $M$ and $Q$ are ADM mass and
charge per unit length in the $z$ direction. This corresponds to the
Kerr-Newman solution in spherical symmetry. The parameter $a$ is
defined in the terms of units of angular momentum per unit mass:

\begin{equation}
\alpha ^{2}a^{2}=1-\frac{\Omega ^{\prime }}{M}.   \label{2.23}
\end{equation}
The relation between $J$ and $a$ is given by
\begin{equation}
J=\frac{3}{2}aM\sqrt{1-\frac{a^{2}\alpha ^{2}}{2}} ,  \label{2.25}
\end{equation}
and the range of $a$ is $0\leq a\alpha \leq 1$. The event horizons
can be found by putting $g^{11}$ equal to zero, giving
\begin{equation}
\alpha ^{2}r^{2}-\frac{4M\left( 1-\frac{3\alpha ^{2}a^{2}}{2}\right)
}{ \alpha r}+\frac{4Q^{2}}{\alpha ^{2}r^{2}}\frac{\left(
1-\frac{3\alpha ^{2}a^{2}}{2}\right) }{\left( 1-\frac{\alpha
^{2}a^{2}}{2}\right) }=0, \label{4.41}
\end{equation}%
which is a quartic equation in $r$. Discarding the two imaginary
roots, the two real roots give the horizons of rotating and charged
black string at
\begin{equation}
r_{\pm }=\frac{b^{\frac{1}{3}}}{2\alpha }\left[ \sqrt{s}\pm
\sqrt{2\sqrt{s^{2}-Q^{2}\left( \frac{2}{M}\right)
^{\frac{4}{3}}}-s}\right] , \label{hor}
\end{equation}
where

\begin{equation}
s=x^{\frac{1}{ 3}}+y^{\frac{1}{3}} , b=4M\left( 1-\frac{3\alpha
^{2}a^{2}}{2}\right).
\end{equation}
We can also write it in a more convenient form as

\begin{equation}
r_{\pm }=\frac{\left[ M\left( 1-\frac{3\alpha ^{2}a^{2}}{2}\right) \right] ^{%
\frac{1}{3}}}{2^{\frac{1}{3}}\alpha }\left[ \sqrt{x^{\frac{1}{
3}}+y^{\frac{1}{3}}}+\sqrt{-x^{\frac{1}{3}
}-y^{\frac{1}{3}}+2\sqrt{-z+\left[ x^{\frac{1}{
3}}+y^{\frac{1}{3}}\right] ^{2}}}\right] , \label{4.55}
\end{equation}
with
\begin{eqnarray*}
x &=&\frac{1}{2}+\frac{1}{2}\sqrt{1-4\left( \frac{4^{\frac{2}{3}}Q^{2}}{%
3\left( 1-\frac{\alpha ^{2}a^{2}}{2}\right) \left( 1-\frac{\alpha ^{2}a^{2}}{%
2}\right) ^{\frac{1}{3}}}\right) ^{3}} {, } \\
 {\ \ }y &=&\frac{1}{2}-\frac{1}{2}\sqrt{1-4\left( \frac{4^{\frac{2}{3}%
}Q^{2}}{3\left( 1-\frac{\alpha ^{2}a^{2}}{2}\right) \left(
1-\frac{\alpha
^{2}a^{2}}{2}\right) ^{\frac{1}{3}}}\right) ^{3}},  \\
{ \ \ }z &=&\frac{4^{\frac{2}{3}}Q^{2}}{\left( 1-\frac{\alpha
^{2}a^{2}}{2}\right) \left( 1-\frac{\alpha ^{2}a^{2}}{2}\right) ^{\frac{1}{3}%
}} .
\end{eqnarray*}
As before we take $r_{\pm}$ as the outer and inner horizons.

In order to calculate the angular velocity of charged and rotating
black string, $\Omega =g_{t\phi }/g_{\phi \phi}$, we note that

\begin{equation}
g_{t\phi }=-\frac{a\sqrt{1-\frac{\alpha ^{2}a^{2}}{2}}}{\left( 1-\frac{%
3\alpha ^{2}a^{2}}{2}\right) } \alpha ^{2}r^{2},  \label{4.56}
\end{equation}%
and

\begin{equation}
g_{\phi \phi }=\frac{r^{2}\left( 1-\frac{\alpha ^{2}a^{2}}{2}\right) }{%
\left( 1-\frac{3\alpha ^{2}a^{2}}{2}\right) }.  \label{4.57}
\end{equation}
Thus

\begin{equation}
\Omega =\frac{a\alpha ^{2}}{1-\frac{\alpha ^{2}a^{2}}{2}}.
\label{4.58}
\end{equation}

To work out the surface gravity of this black string, let us
consider the general form of rotating cylindrical metric as
\begin{equation}
ds^{2}=-g_{tt}dt^{2}+g_{rr}dr^{2}+g_{\phi \phi }d\phi
^{2}+g_{zz}dz^{2}-2g_{t\phi }dtd\phi .  \label{4.16}
\end{equation}
To get rid of the motion of $\phi $ we employ the following rotating
coordinate system that is co-rotating with the horizon
\begin{equation}
\phi =\phi ^{\prime }+\Omega  t, { \ \ \ \ \ \ \ }\phi =\phi
^{\prime }-\Omega  t .   \label{4.17}
\end{equation}
Using this transformation the above metric becomes

\begin{equation}
ds^{2}=-G_{tt}dt^{2}+g_{rr}dr^{2}+g_{zz}dz^{2}+g_{\phi \phi }d\phi
^{\prime 2}-2g_{t\phi }^{\prime }dtd\phi ^{\prime }.  \label{4.19}
\end{equation}
Here
\begin{equation}
G_{tt}=g_{tt}+2g_{t\phi }\Omega -g_{\phi \phi }\Omega ^{2}.
\label{4.20}
\end{equation}
This metric has a Killing field
\begin{equation}
\xi ^{\mu }=\frac{\partial }{\partial t}+\Omega \frac{\partial
}{\partial \phi },  \label{4.22}
\end{equation}
which satisfies
\begin{equation}
\xi ^{\mu }\xi _{\mu }=g_{tt}+2g_{t\varphi }\Omega -g_{\phi \phi
}\Omega ^{2}.  \label{4.23}
\end{equation}%
Comparing Eqs. (\ref{4.20}) and (\ref{4.23}) we see that $\xi ^{\mu
}\xi _{\mu }=G_{tt}$. Writing  \cite{21, 5} $\xi ^{\mu }\xi _{\mu
}=-\lambda ^{2}$, we note that $G_{tt}=-\lambda ^{2}$. Therefore we
get
\begin{eqnarray}
\nabla _{\mu }\left( -\lambda ^{2}\right) &=&-2\kappa \xi _{\mu } ,
\nonumber  \\
g^{a\mu } \nabla _{\mu }\left( -\lambda ^{2}\right)  \nabla
_{\mu }\left( -\lambda ^{2}\right) &=&4\kappa ^{2}G_{tt} ,  \nonumber  \\
g^{rr}\left( \partial _{r}G_{tt}\right) ^{2} &=&4\kappa ^{2}G_{tt} .
\end{eqnarray}
Thus
\begin{equation}
\kappa =\frac{1}{2}\sqrt{g^{rr}} \frac{\partial _{r}\left(
G_{tt}\right) }{\sqrt{G_{tt}}} .  \label{4.24}
\end{equation}
Now using Taylor's series expansions
\begin{eqnarray}
G_{tt} &=&G_{tt}^{\prime }\left( r-r_{o}\right) +...,  \label{4.25} \\
g^{rr} &=&g^{rr\prime }\left( r-r_{o}\right) +...,  \label{4.26}
\end{eqnarray}
in the above formula, the surface gravity of rotating black string
becomes

\begin{equation}
\kappa =\frac{\sqrt{G_{tt}^{\prime } g^{rr\prime }}}{2} .
\end{equation}
Using the values

\begin{eqnarray}
G_{tt} &=&\frac{1-\frac{1}{2}\alpha ^{2}a^{2}}{1-\frac{3}{2}\alpha ^{2}a^{2}}%
\left[ \alpha ^{2}r^{2}-\frac{4M\left( 1-\frac{3}{2}\alpha
^{2}a^{2}\right) }{\alpha r}+\frac{4Q^{2}}{\alpha ^{2}r^{2}}\left(
\frac{1-\frac{3}{2}\alpha ^{2}a^{2}}{1-\frac{1}{2}\alpha
^{2}a^{2}}\right) \right] ,  \label{4.72}
\end{eqnarray}%
and
\begin{equation}
g^{rr}=\alpha ^{2}r^{2}-\frac{4M\left( 1-\frac{3}{2}\alpha
^{2}a^{2}\right)
}{\alpha ^{2}r}+\frac{4Q^{2}}{\alpha ^{2}r^{2}}\left( \frac{1-\frac{3}{2}%
\alpha ^{2}a^{2}}{1-\frac{1}{2}\alpha ^{2}a^{2}}\right)
,\label{4.73}
\end{equation}%
this takes the form
\begin{equation}
\kappa =\sqrt{\frac{1-\frac{1}{2}\alpha
^{2}a^{2}}{1-\frac{3}{2}\alpha ^{2}a^{2}}}\left[ \alpha
^{2}r_{+}+\frac{2M\left( 1-\frac{3}{2}\alpha
^{2}a^{2}\right) }{\alpha ^{2}r_{+}^{2}}-\frac{4Q^{2}}{\alpha ^{2}r_{+}^{3}}%
\left( \frac{1-\frac{3}{2}\alpha ^{2}a^{2}}{1-\frac{1}{2}\alpha ^{2}a^{2}}%
\right) \right] .  \label{4.74}
\end{equation}%
On using Eq. (\ref{4.55}) in this expression we obtain after
simplification
\begin{equation}
\kappa =\frac{\alpha M^{\frac{1}{3}}\sqrt{1-\frac{\alpha ^{2}a^{2}}{2}}D}{2^{%
\frac{1}{3}}\left( 1-\frac{3\alpha ^{2}a^{2}}{2}\right) ^{\frac{1}{6}}}+%
\frac{2^{\frac{5}{3}}M^{\frac{1}{3}}\sqrt{\left( 1-\frac{1}{2}\alpha
^{2}a^{2}\right) }}{\left( 1-\frac{3}{2}\alpha ^{2}a^{2}\right) ^{\frac{1}{6}%
}D^{2}}-\frac{8\alpha Q^{2}}{MD^{3}\sqrt{\left( 1-\frac{\alpha ^{2}a^{2}}{2}%
\right) \left( 1-\frac{3\alpha ^{2}a^{2}}{2}\right) }}.
\label{4.75}
\end{equation}
Here
\begin{equation}
D=\sqrt{x^{\frac{1}{3}}+y^{\frac{1}{3}}}
+\sqrt{-x^{\frac{1}{3}}-y^{\frac{1}{3}}+2\sqrt{ -z+\left(
x^{\frac{1}{3}}+y^{\frac{1}{3}} \right) ^{2}}} . \label{4.76}
\end{equation}
Thus the Hawking temperature for the charged rotating black string
is given by

\begin{equation}
T=\frac{1}{2\pi }\sqrt{\frac{1-\frac{1}{2}\alpha
^{2}a^{2}}{1-\frac{3}{2} \alpha ^{2}a^{2}}}\left[ \alpha
^{2}r_{+}+\frac{2M\left( 1-\frac{3}{2}\alpha ^{2}a^{2}\right)
}{\alpha ^{2}r_{+}^{2}}-\frac{4Q^{2}}{\alpha ^{2}r_{+}^{3}} \left(
\frac{1-\frac{3}{2}\alpha ^{2}a^{2}}{1-\frac{1}{2}\alpha ^{2}a^{2}}
\right) \right] .  \label{4.77}
\end{equation}
In terms of $D$ this can be written as
\begin{equation}
T=\frac{1}{2\pi }\left[ \frac{\alpha
M^{\frac{1}{3}}\sqrt{1-\frac{\alpha
^{2}a^{2}}{2}}D}{2^{\frac{1}{3}}\left( 1-\frac{3\alpha
^{2}a^{2}}{2}\right)
^{\frac{1}{6}}}+\frac{2^{\frac{5}{3}}M^{\frac{1}{3}}\sqrt{\left( 1-\frac{1}{2%
}\alpha ^{2}a^{2}\right) }}{\left( 1-\frac{3}{2}\alpha ^{2}a^{2}\right) ^{%
\frac{1}{6}}D^{2}}-\frac{8\alpha Q^{2}}{MD^{3}\sqrt{\left(
1-\frac{\alpha ^{2}a^{2}}{2}\right) \left( 1-\frac{3\alpha
^{2}a^{2}}{2}\right) }}\right] . \label{4.78}
\end{equation}

The area of the horizon per unit length of charged rotating black
string takes the form

\begin{equation}
\sigma =\frac{2^{\frac{1}{3}}\pi \left[ M\left( 1-\frac{3\alpha ^{2}a^{2}}{2}%
\right) \right] ^{\frac{2}{3}}D^{2}}{\alpha },   \label{4.79}
\end{equation}
and the entropy in term of the horizon area becomes

\begin{equation}
S=\frac{\pi \left[ M\left( 1-\frac{3\alpha ^{2}a^{2}}{2}\right) \right] ^{%
\frac{2}{3}}D^{2}}{2^{\frac{5}{3}}\alpha }.  \label{4.80}
\end{equation}
The electric potential in this case takes the form

\begin{equation}
\Phi =\frac{2^{\frac{4}{3}}Q}{\left[ M\left( 1-\frac{3\alpha ^{2}a^{2}}{2}%
\right) \right] ^{\frac{1}{3}}D} .   \label{4.82}
\end{equation}

\section{Discussion}

We have studied thermodynamics of uncharged, charged and rotating
black strings having a negative cosmological constant. These black
configurations are asymptotically anti-de Sitter. We note that
thermodynamic behaviour of these objects is quite different from
their corresponding spherical configurations like the Schwarzchild,
Kerr and Kerr-Newman black holes. This shows that thermodynamic
properties change with the change in topological structure.  The
negative cosmological constant also has a significant role here.

We have used two different methods to work out the surface gravity
of non-rotating and rotating strings, and the results are found to
be consistent. For rotating and charged black string we have used
the Killing field $\xi ^{\mu }= \partial /\partial t+\Omega
\partial / \partial \phi $, where $\partial /\partial
t$ is the time like Killing vector, $\partial /\partial \phi$ is the
rotational Killing vector and $\Omega $ is the angular velocity. In
the most general case of charged and rotating black string the
Hawking temperature is found to be
\begin{equation}
T=\frac{1}{2\pi }\sqrt{\frac{1-\frac{1}{2}\alpha ^{2}a^{2}}{1-\frac{3}{2}%
\alpha ^{2}a^{2}}}\left[ \alpha ^{2}r_{+}+\frac{2M\left(
1-\frac{3}{2}\alpha
^{2}a^{2}\right) }{\alpha ^{2}r_{+}^{2}}-\frac{4Q^{2}}{\alpha ^{2}r_{+}^{3}}%
\left( \frac{1-\frac{3}{2}\alpha ^{2}a^{2}}{1-\frac{1}{2}\alpha ^{2}a^{2}}%
\right) \right] .  \label{5.3}
\end{equation}
For uncharged rotating black string, i.e. $Q=0$, it reduces to
\begin{equation}
T=\frac{1}{2\pi }\sqrt{\frac{1-\frac{1}{2}\alpha ^{2}a^{2}}{1-\frac{3}{2}%
\alpha ^{2}a^{2}}}\left[ \alpha ^{2}r_{+}+\frac{2M\left(
1-\frac{3}{2}\alpha ^{2}a^{2}\right) }{\alpha ^{2}r_{+}^{2}}\right]
.  \label{5.4}
\end{equation}
For charged black string without rotation, i.e. we put $a=0$, the
Hawking temperature becomes
\begin{equation}
T=\frac{1}{2\pi }\left( \alpha ^{2}r_{+}+\frac{2M}{\alpha r_{+}^{2}}-\frac{%
4Q^{2}}{\alpha ^{2}r_{+}^{3}}\right) ,  \label{5.5}
\end{equation}
which is the same as in \cite{CZ} and has been confirmed by the
Hamilton-Jacobi method in the quantum tunneling approach
\cite{AS11a}. For the simplest case when $Q=0=a$ the temperature
takes the form
\begin{equation}
T=\frac{1}{2\pi }\left( \alpha ^{2}r_{+}+\frac{2M}{\alpha
r_{+}^{2}}\right) . \label{5.6}
\end{equation}

It must be mentioned here that in the case of uncharged rotating
black string the horizon can be written from Eq. (\ref{hor}) by
taking $Q=0$ as

\begin{equation}
r_{+}^3=\frac{2}{\alpha^3}\left[-M+3\sqrt {M^2-\frac{8}{9}\alpha^2
J^2} \right] , \label{hor1}
\end{equation}
which is different from the value given in Ref. \cite{lemos95b} by a
factor of $8$. From the horizon given in Eq. (\ref{hor1}) we can
correctly obtain the value for the case of uncharged non-rotating
black string. However, the horizons given in Ref. \cite{LZ} are
correct and reduce to the uncharged case by putting $Q=0$. The
Hawking temperature is not given in Ref. \cite{LZ}, and is
calculated in Ref. \cite{lemos95b} for the uncharged rotating black
string, but it is not correct. If we use the notation of Ref.
\cite{lemos95b} i.e. we write Eq. (\ref{5.4}) in terms of $J$ by
using  Eq. (\ref{2.25}) then the formula comes out to be

\begin{equation}
T=\frac{1}{2^{\frac{2}{3}} \alpha \pi}\left(3\sqrt
{M^2-\frac{8}{9}\alpha^2 J^2}-M \right)^{\frac{1}{3}} \left(\sqrt
{\frac{1-\sqrt{1-\frac{8}{9}\frac{\alpha^2
J^2}{M^2}}}{3\sqrt{1-\frac{8}{9}\frac{\alpha^2 J^2}{M^2}}
-1}}+1\right), \label{temp}
\end{equation}
which is different from the one given in Ref. \cite{lemos95b}.



\end{document}